\documentclass{ws-p8-50x6-00}

\begin{document}

\title{Cosmological magnetic fields by parametric resonance?}

\author{Fabio Finelli}

\address{Dept. of Physics, Purdue University, West
Lafayette, IN 47907, USA\\
Dip. di Fisica, Univ. degli Studi di Bologna
and INFN, 40126 Bologna, Italy 
\\E-mail: fabio@physics.purdue.edu}


\maketitle

\abstracts{
We investigate the possibility that electromagnetic fluctuations are
amplified in expanding universe by parametric resonance, during the
oscillatory regime of a scalar field to which they are coupled.
We consider scalar electrodynamics and we find that electromagnetic
fluctuations undergo exponential instabilities. 
This mechanism could have some
relevance for the problem of large scale primordial
magnetic fields.}

\section{Introduction}
The origin of $\mu$G galactic and intergalactic fields
which are coherent over Kpc scales
is still a puzzle for astrophysicists \cite{review}. These could be
generated during
galaxy formation or just amplified by a dynamo effect from primordial
seeds of cosmological origin \cite{review2} (see also O. Tornkvist in
this volume).

The most important requirements to get reasonably agreement with
observations concern the amplitude and the large coherence length of
the magnetic fields observed. 
The amplitude of the magnetic field redshifts due to the flux
conservation. Moreover, there is a causal upper bound to the scale 
of coherence at which the
magnetic seed is generated, related to the horizon scale (and then to the
Hubble scale $H^{-1}$ if there is not a mismatch between these two
scales).

This latter condition is the most serious problem for all the 
{\em microphysical} mechanisms, such as thermal  
fluctuations in plasma \cite{plasma}, phase transition \cite{pt}, just
to cite few examples, and it
resembles the large scale structure (LSS) problem for the Big Bang
scenarios.
The mechanisms produce easily a magnetic seed 
with an amplitude of the right order of magnitude, but on too small scales.

In a totally different perspective, one can try to address 
the problem of cosmological magnetic fields as the LSS problem 
was faced with {\em inflation}.
Then the scale is no more a problem, but the amplitude 
becomes a serious issue. This fact is due to the conformal invariance of the 
electromagnetic field in conformally invariant space-times, such as 
the cosmological ones. For this reason electromagnetic fields do not 
suffer amplification by the time-dependent geometry, as minimally coupled 
scalar fields, gravitational waves do. Therefore, 
the breaking of the conformal invariance was investigated
as one of the necessary conditions for the generation of
primordial magnetic fields by Turner and Widrow \cite{turner}.
By using alternative theories of gravity (where the
electromagnetic field is not conformally coupled to a conformally
flat geometry) magnetic seeds of the right order of magnitude can be
achieved \cite{ratra}. 

When the electromagnetic (EM) field is considered in 
interaction with other fields besides gravity -as with a charged 
scalar field for instance-, then 
its conformal invariant property is naturally broken 
because of the interaction term. The study of interacting quantum 
fields in cosmological spacetimes has received a renewed interest 
since it was realized that parametric resonance (PR) 
could play an important role \cite{para}. In the context of generation of
cosmological magnetic fields 
a little attention was paid to the possibility that PR
could play some effect in the interaction of the EM field
with other fields. Even during a phase transition,  
the coherent oscillation of the order parameter around
its minimum could enhance the production of fields coupled to it. 
In our opinion, there are four reasons in order to consider  
the effect of PR for EM fields 

a) the photon is massless, and this is almost a necessary condition in
order 
to have appreciable amplification in expanding universe without
considering very large couplings \cite{klsbig}

b) the effect due to PR for fields which would be conformally invariant 
without the interaction is less disputed by the expansion of the
universe, because the equations of motion are conformally related 
to equations of motion in the Minkowsky spacetime \cite{greene}

c) the growth due to parametric resonance is exponential as 
the one predicted by the dynamo effect \cite{dynamo}, one of
the astrophysical processes postulated to explain the observational evidence
for galactic magnetic fields

d) there are several examples in which the effect of PR is manifest on
the maximum causal scale allowed by the problem (the coherence scale of the
field). Since the scale of the magnetic seeds is always a 
crucial issue for a model which aims to explain their origin, this feature 
of PR is potentially very interesting.

\section{A scalar electrodynamics model}
We consider the Abelian-Higgs model in Robertson-Walker background
\cite{ours1}
\be
{\cal L} = - \frac{1}{16\pi} F_{\mu \nu} F^{\mu \nu} - (D_\mu \Phi)^*
(D^\mu \Phi) - V(\Phi^* \Phi)
\label{scalarlag}
\ee     
By working in the Coulomb gauge, the equation of motion for Fourier 
component of the transverse fluctuations of the gauge field are 
\be
{\bf A}''_{T k} + (k^2 + 4 \pi e^2 a^2 \rho^2) {\bf A}_{T k} = 0 \,,
\label{main}
\ee
where a prime denotes derivative with respect to the conformal time
$\eta$ and $\rho$ is implicitly defined as $\Phi (t) = e^{i \Theta(t)}
\rho(t)/\sqrt{2}$ and satisfies   
\begin{equation}
\ddot \rho + 3 H \dot \rho + \frac{\partial
V}{\partial \rho} = 0 \,,
\label{homfield}
\end{equation}
where we have neglected any back-reaction term.
The non-dynamical quantity $A_{0 k}$ is given by the 
equation analogous to the Gauss law.

Equation (\ref{main}) describes a harmonic oscillator with time
dependent frequency: during the oscillation of the complex scalar field
Eq. (\ref{main}) can be reduced to a Mathieu-like equation.
The solutions to this type of equation show an exponential instability
$\propto e^{\mu_k \eta}$, for some interval
of frequencies, called {\em resonance bands}. 
The structure of the resonance bands 
depends strongly on the time behaviour of the homogeneous scalar field,
which on turn depends on the form of the potential $V(\Phi)$. An example 
of stable resonance on long wavelengths is given by 
$V = \lambda (\Phi^* \Phi)^2$ with $4 \pi e^2 = 2 \lambda$.
The physical fields and the energy density $T_{00}^{\mathrm{EM}}$ are
obtained through 
\be
{\bf E} = \frac{1}{a}(\nabla{A_0} - {\bf A}') \quad {\bf B} =
\frac{1}{a} \nabla \times {\bf A} \quad T_{00}^{\mathrm{EM}} = 
\frac{1}{8 \pi a^2} \left( {\bf E}^2 + {\bf B}^2 \right)
\ee

\section{Discussions and Conclusions}

The influence of plasma effects on the amplification driven by a coherent
condensate is discussed in Ref. [11]: generically, the presence of
other not coherent charges counteracts the resonance. However, one can
think that in 
preheating after inflation, when it is assumed that the "creation" of
charged particles is contemporaneous to the amplification of EM
fluctuations, the resonance could proceed just as worked out in the vacuum
case. In this way a charged inflaton could play the role of the condensate. 
If the resonance occurs on long wavelengths, then this would be on
observable scales, because of the coherence of the inflaton on
super-Hubble scales [12].  

This would lead to an additional growth of the fraction of the
electromagnetic energy density relative to the total one during
reheating, which is missed in the usual
predictions based on inflation-inspired models:
\begin{displaymath}
r_{\mathrm{reh}} \equiv \left. \frac{\rho_\gamma}{\rho_{\mathrm{tot}}}
\right|_{\mathrm{reh}} =
r_{\mathrm{infl}} \times \left\{
\begin{array}{ccc} 
e^{2 \mu_k \eta}/\eta^2  & \quad \mathrm{for} & \quad V = m^2 \Phi^* \Phi
\\
e^{2 \mu_k \eta}  \quad & \mathrm{for} & \quad V = \lambda (\Phi^* \Phi)^2
\end{array} \right.
\end{displaymath} 

The interpretation of this growth of electromagnetic fluctuations as 
magnetic fields relies on the same idea used for inflation and
string-produced magnetic fields \cite{turner,ratra}: on large scale the
large conductivity suppresses the electric part, and the magnetic part
evolves along with magnetic flux conservation.

\section*{Acknowledgments}
This work is in collaboration with A. Gruppuso. 
We thank the organizer of COSMO-99 for a stimulating and 
enjoyable conference and R. Brandenberger, A.-C. Davis, K. Dimopoulos, 
A. Riotto, L. Sorbo, O. Tornkvist comments, discussions and questions.

\end{document}